\documentclass[12pt,twoside]{article}
\usepackage{mattstyle}  % Loads my formatting
\usepackage[utf8]{inputenc}
\graphicspath{ {figures/} }

\usepackage[flushleft]{threeparttable} % Table notes

\makeatletter
\def\thanks#1{\protected@xdef\@thanks{\@thanks
        \protect\footnotetext{#1}}}
\makeatother

\usepackage[section]{placeins}

\usepackage[super]{nth}

\theoremstyle{definition}
\newtheorem{definition}{Definition}

\usepackage[small]{titlesec}

\begin{document}

%%%%%%%%%%%%%%%%%%%%%%%%%%%%%%%%%%%%%%%%%%%%%%%%%%%%%%%%%%%%%%%%%
% TITLE PAGE
%%%%%%%%%%%%%%%%%%%%%%%%%%%%%%%%%%%%%%%%%%%%%%%%%%%%%%%%%%%%%%%%%

%\begin{singlespacing}

\begin{titlepage}

\begin{spacing}{0.9}

\author{Francis Bloch\\{\small Paris School of Economics}
\and Matthew Olckers\\{\small Monash University}
}

\title{Friend-Based Ranking in Practice}
\runningheads{Bloch \& Olckers}{Friend-Based Ranking in Practice}

\date{\small \today}
\maketitle
\thispagestyle{empty}

%\clearpage
\vspace{-0.3in}

\end{spacing}

\begin{abstract}
A planner aims to target individuals who exceed a threshold in a characteristic, such as wealth or ability.  The individuals can rank their friends according to the characteristic. We study a strategy-proof mechanism for the planner to use the rankings for targeting. We discuss how the mechanism works in practice, when the rankings may contain errors.
\end{abstract}

\end{titlepage}

%------------------------------------------------------------------------

\section{Introduction}

Through social interactions we learn a great deal about our friends, including characteristics such as wealth and ability. Although we may not be able to put a precise number to these characteristics, we can rank our friends with ease---from poorest to richest, from least to most business savvy, or from shy to outgoing. 

These rankings can be extremely valuable for planners, such as government agencies and banks. For example, a bank may want to know which entrepreneur is most profitable, or a government agency may want to know which household is in the most need.

In \citet{FBR2020} we study how a planner can design a mechanism to extract a ranking of a unidimensional characteristic from a network of individuals. We assume that each individual only observes an ordinal ranking of the characteristics of their friends---their neighbors in the network. We show that a  planner can design a mechanism to extract all available information if and only if every pair of friends has a friend in common. In many social networks this property is satisfied---most friends have many friends in common.

In \citet{FBR2020} we made the assumption that individuals have perfect ordinal information about their friends. Individuals do not mistakenly rank friends in the incorrect order. This assumption is unlikely to be satisfied in practice. The observed rankings may be perturbed, especially when friends are very similar. The mechanism we derived relied on the assumption of perfect information. If the planner observed conflicting reports, she could conclude that at least one of the individuals was lying. 

In this article, we extend the model of \citet{FBR2020} by allowing individuals to make errors when observing the ranking of their friends. We introduce a mechanism to target a group of agents rather than attempting to elicit a complete ranking. To aggregate possibly conflicting and incomplete ranks, we rely on the HodgeRank algorithm, which has the advantage of including a goodness-of-fit measure \citep{jiang2011hodgerank}. Finally, we include a brief empirical exercise to compare our friend-based ranking mechanism with the traditional approach of community-based targeting.

We hope that friend-based ranking mechanisms can be used in practice. The article is accompanied by well-documented computer code (available at \texttt {\href{https://github.com/matthewolckers/fbr-in-practice/}{github.com/matthewolckers/fbr-in-practice}}) so that practitioners can apply our methods to their ranking data.

\section{Model}

\subsection{Planner Targets Individuals}

We consider a community $N$ of individuals $i=1,2,..n$ with characteristic $\theta_i \in \mathbb{R}$. The individuals belong to a social network $g$. The planner designs a mechanism to select a targeted group of individuals, denoted by the set $T \subset N$ according to their characteristic $\theta_i$.

The characteristics $\theta_i$ are not directly observed by the planner who relies on the reports $r^i$ sent by individuals. Individuals have local, ordinal information: their types are ranks of their friends. As opposed to \citet{FBR2020}, we do not assume that the rankings are perfect. Instead, individuals make mistakes and may observe an incorrect ranking of their friends. We do not make any specific assumption on the information structure generating an individual's ranking as a function of the true characteristics of his friends. 

The objective of the planner is to maximize the number of individuals in $T$ for which $\theta_i>\underline{\theta}$ and minimize the number of agents in $T$ for which $\theta_i < \underline{\theta}$, where $\underline{\theta}$ is a fixed threshold. Each individual obtains a payoff of $1$ if he is chosen in the targeted set $T$ and $0$ otherwise.

\subsection{Mechanism}

Building on \citet{mattei2020},  we construct the following three-step mechanism ${\cal T}$ assigning a subset $T$ of $N$ to every vector of individual reports ${\bf r} = (r^1,...,r^n)$.
\vspace{2mm}

\noindent \emph{Step 1 - Collect rankings}

Each individual reports a ranking of his friends in the social network.
\vspace{2mm}

\noindent \emph{Step 2 - Estimate scores}

The planner uses an algorithm to map the reports ${\bf r}$ to a score $s_i \in \mathbb{R}$ for each individual $i$. To prevent manipulation, $i$'s report is excluded from the calculation of his score, i.e. $s_i({\bf r}) = s_i({\bf r'})$ for any ${\bf r}, {\bf r'}$ such that $r_{-i} = r'_{-i}$.
\vspace{2mm}

\noindent \emph{Step 3 - Choose the targeted group $T$}

If $s_i > \underline{s}$, a predetermined cutoff, individual $i$ is included in $T$. If $s_i \leq \underline{s}$, $i$ is excluded from $T$.

%--------------------------------------------------------------------
\section{Properties of the Mechanism}

We first recall that a mechanism is strategy-proof if no individual has an incentive to misreport and group-strategy-proof if no subset of individuals has an incentive to misreport.

\begin{definition} A mechanism  is strategy-proof if for every $i$, every type $r_i$ of $i$, every report $r_{-i}$ of the other individuals,
\[ \Pr(i \in T(r_i,r_{-i}) \geq \Pr(i \in T(r'_i,r_{-i})\]
\noindent for any other report $r'_i$ of individual $i$.
\end{definition}

\begin{proposition} \label{pro:strategyproof} The three-step mechanism ${\cal T}$ is strategy-proof.
\end{proposition}

The proof of Proposition \ref{pro:strategyproof} is immediate. Because the score of individual $i$, $s_i$, is independent of $i$'s report, and individual $i$ is only included in the target set $T$ if and only if $s_i > \underline{s}$, $i$ cannot change her probability of inclusion by changing her report. Hence, the mechanism is strategy-proof.

\bigskip

We say that the scoring function $s({\bf r})$ is monotonic if $s_i({\bf r}) > s_i({\bf r'})$ whenever ${\bf r}$ ranks player $i$ higher than ${\bf r'}$.\footnote{We say that ${\bf r}$ ranks player $i$ higher than ${\bf r'}$ if,  whenever $i$ beats $j$ under ${\bf r'}$, $i$ beats $j$ under ${\bf r}$ and there is an instance where $i$ beats $j$ under ${\bf r}$ but not under ${\bf r'}$.} When the scoring rule is monotonic, the three-step mechanism reacts to changes in the reports of the individuals. We show however that it is not immune to deviations by groups of players.

\begin{definition} A mechanism  is manipulable by a coalition $S$ at the report ${\bf r}$ if there exists a vector of reports $(r'_j)_{j \in S}$ such that, for every individual $i \in S$, 
\[ Pr (i \in T((r'_j)_{j \in S}, (r_k)_{k \in S}) \geq Pr (i \in T({\bf r})),\] 
\noindent with strict inequality for some individual.
A mechanism ${\cal T}$ is group-strategy proof if it is not group manipulable by a coalition $S$ at any type profile ${\bf r}$.
\end{definition}

\begin{proposition} \label{pro:groupstrategyproof}  Suppose that $n \geq 3$. If the scoring rule is monotonic, there exists a threshold level $\underline{s}$ such that the three-step mechanism is not group-strategy-proof.
\end{proposition}

The proof of Proposition \ref{pro:groupstrategyproof} only requires a counter-example. Consider two connected individuals $i$ and $j$. Let $k$ be an individual connected to $j$.   Suppose that $j$ observes $\theta_k > \theta_i$,  and consider a profile ${\bf r}$ such that $s_i({\bf r}) > s_j({\bf r}).$ \footnote{If no such profile exists, we revert the roles of $i$ and $j$ in the example.}  Because the scoring rule is monotonic, if $j$ reports $\theta_i > \theta_k$, the score of individual $i$ is higher under the new vector of reports ${\bf r'}$,  $s_i ({\bf r'}) > s_i({\bf r})$. Now choose $\underline{s} \in (s_i({\bf r}), s_i({\bf r'}))$. The three-step mechanism is group manipulable by the coalition $(\{i,j\})$ at the profile ${\bf r}$.

\bigskip

Finally, we note that the threshold for inclusion $\underline{s}$ must be an absolute, rather than a relative value. Consider a variation of the mechanism, where the size of the targeted group is fixed, and agents are included according to their relative score. Formally, let the fraction of targeted individuals be equal to  $\alpha = \frac{|T|}{n}$  and let individuals be assigned to $T$ if and only if 
\begin{align*}
    |\{j, s_j > s_i\}| \geq n(1-\alpha).
\end{align*} 
Consider a community of three individuals, $i,j,k$ connected in a triangle, and let $\alpha = \frac{1}{3}$. Consider a scoring rule which is monotonic and anonymous. A scoring rule $s$ is anonymous if, for any permutation of the individuals $\pi$, for any player $i$, $s_{\pi(i)}(\pi({\bf r})) = s_i({\bf r})$.

Let the report ${\bf r}$ be such that individual $i$ ranks $j$ above $k$, individual $j$ ranks $i$ above $k$ and individual  $k$ ranks $j$ above $i$. By anonymity, and because all vertices are similar in a triangle, the scoring rule only depends on the number of times that an individual beats another individual. Hence individual $j$ is chosen with probability $1$ after the report ${\bf r}$. If however $i$ changes here report and ranks $k$ above $j$, all the individuals have an equal chance to be included in the target set.

%--------------------------------------------------------------------
\section{Aggregating Reports using HodgeRank}

The planner must choose an algorithm to map the reported rankings into the scores, $s_i$. We propose using HodgeRank, an approach that can handle cycles and incomplete ranking data \citep{jiang2011hodgerank}. Also, the HodgeRank method does not rely on any assumption on the distribution of errors, and hence can be applied when the planner does not have any information on the community. In this section, we explain the HodgeRank algorithm.

We start with reported ranks of the form $r_{i j}^k \in \{-1,1\}$ where $r_{i j}^k=1$ if $k$ ranks $i$ above $j$ and $r_{i j}^k=-1$ if $k$ ranks $i$ below $j$. We collect the reported ranks into a weighted and directed ranking graph $Y$ with edge set $E$. Each node corresponds to an individual. The weighted edges $Y_{ij}$ aggregate the reported ranks about each pair of individuals. If there are multiple reported ranks on a specific pair, we take the mean of the reported ranks.

HodgeRank chooses a score $s_i$ for each individual $i$ to minimize the squared difference between the scores and the aggregated rankings. HodgeRank solves the problem:
\begin{align*}
    \min\left[\sum_{\{i, j\} \in E} \left((s_i - s_j) - Y_{i j}\right)^{2}\right]
\end{align*}
\citet{jiang2011hodgerank} showed that the residual of this least squares problem corresponds to the cycles in the ranking graph. The \textit{cycle ratio} provides a measure of goodness-of-fit for the estimated scores.
\begin{align*}
    \text{Cycle ratio} = \dfrac{\sum_{\{i, j\} \in E} \left( (\hat{s}_i-\hat{s}_j) - Y_{i j}\right)^{2}}{\sum_{\{i, j\} \in E} (Y_{i j})^2}
\end{align*}
We now provide several examples to illustrate how HodgeRank estimates scores. Let us start with a simple case with four individuals. %
\begin{center}
    \includegraphics{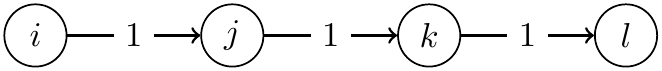}
\end{center} %
From the graph, it is clear the rank has no cycles and the scores $s_i = -1.5, s_j = -0.5, s_k=0.5, s_l=1.5$ solve the least squares problem.\footnote{The scores are unique up to an additive constant. We set the scores to sum to zero.} The cycle ratio is zero as the scores explain all of the variation in the ranking graph.

Let's add a cycle of length three to the graph. % 
\begin{center}
    \includegraphics{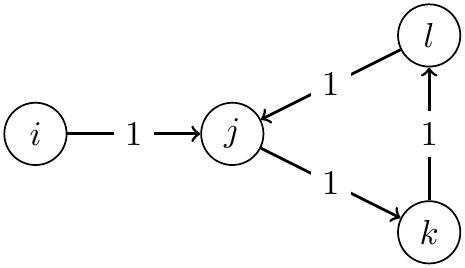}
\end{center} %
\noindent The scores become $s_i = -0.75, s_j = 0.25, s_k=0.25, s_l=0.25$. The cycle ratio is 0.75. For each pair on the cycle, the estimated scores predict no difference on the ranking graph, but these pairs have a difference of 1 unit.

Now let's add a cycle of length four to the first graph. 
\begin{center}
    \includegraphics{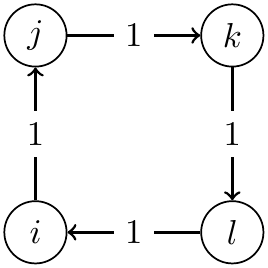}
\end{center}
\noindent The scores become $s_i = 0, s_j = 0, s_k=0, s_l=0$. The cycle ratio is equal to one as the cycle explains all of the variation in the ranks.

The cutoff score \underline{s} requires a minimal level of consensus in the ranking graph for individuals to be included in the targeted set $T$. If $\underline{s}=1$, individual $l$ would be targeted in the first example whereas the targeted set would be empty for the examples with cycles. 

Finally, we consider an example where the cycle ratio may be positive due to the magnitude of the edge weights of the ranking graph. In the example below, the scores are $s_i=-\frac{2}{3}, s_j=0, s_k=\frac{2}{3}$ and the cycle ratio is $\frac{1}{9}$.
\begin{center}
    \includegraphics{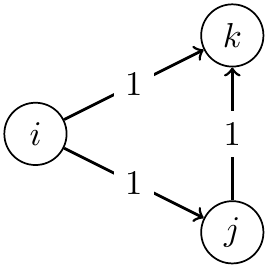}
\end{center}
Note that the HodgeRank algorithm assumes the pairwise comparisons are cardinal. The path $(i,j,k)$, composed of $(i,j)=1$ and $(j,k)=1$, implies that the pairwise comparison $(i,k)=2$. However, in this example $(i,k)=1$. We do not find the same magnitude by taking two different paths between $i$ and $k$. Thus, the discrepancy is reflected in the scores and the cycle ratio. 

Recall that the edges $Y_{ij}$ are formed by taking an average over all individuals who rank the pair $(i,j)$. Therefore, $Y_{ij}$ can range from $-1$ to $1$. Given this restriction on $Y_{ij}$, we are likely to find discrepancies by following different paths along the ranking graph. Other methods of aggregating the reported ranks $r^k_{ij}$ into the weighted edges $Y_{ij}$ may lead to more accurate scores. We leave this problem to future research.

%---------------------------------------------------------------------------
\section{Comparison to Traditional Community-Based Targeting}

One application of friend-based ranking is poverty targeting. In a traditional community-based targeting approach, a community meets together and agrees on a ranking of households from poorest to richest. Friend-based ranking does not require a centralized meeting. Rankings can be collected from each individual separately.

Using data from 423 Indonesian hamlets collected by \citet{alatas2012targeting,alatas2016network}, we can compare the community-based targeting and friend-based ranking approaches. The data contains four essential ingredients: ground-truth data on wealth as measured by a detailed consumption survey, social network data, the outcome of a community-based targeting exercise, and individual rankings of wealth reported by nine households.

Although this data contains all the ingredients required to compare friend-based ranking and community-based targeting, the setting is not ideal since there are only nine households who reported rankings. We anticipate that friend-based ranking will be of greater benefit in much larger networks.

The individual rankings required each of the nine households to rank the other eight from poorest to richest. The respondent was allowed to indicate that he or she did not know the ranking of some individuals. In a brief empirical exercise, we compare the distribution of consumption for the targeted and excluded individuals for each of the two methods: friend-based ranking and community-based targeting.

\begin{figure}
    \centering
    \caption{Comparing Targeting Methods}
    \includegraphics[width=10cm]{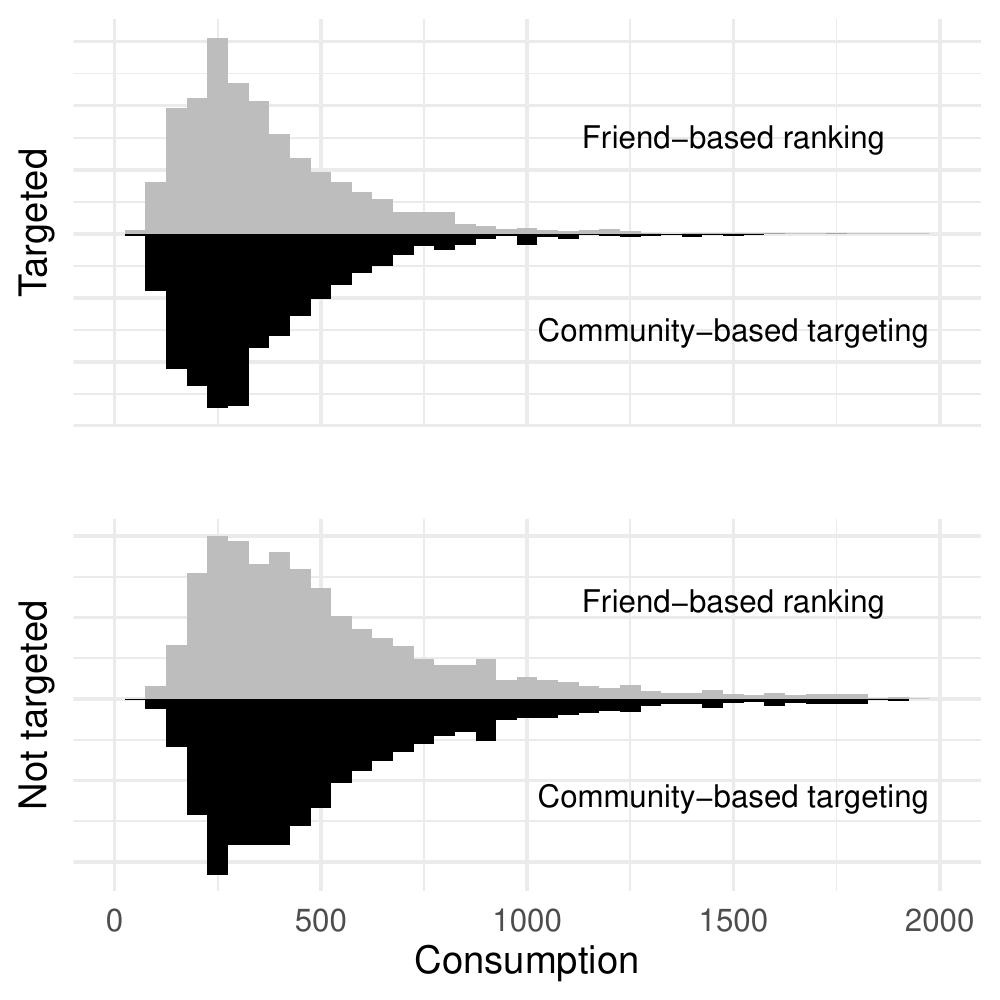}
    \label{fig:1}
    \fignote{4.5in}{Note: Sample size of 3\,522 individuals. The targeted set contains 1\,084 individuals for both methods. For community-based targeting, we use the targeting quota specified for each hamlet. For friend-based ranking, we set the cutoff score $\underline{s}$ to include a similar number of individuals in the targeted group. This cutoff is fixed and does not vary between hamlets.}
\end{figure}

Figure \ref{fig:1} shows the distribution of consumption for households targeted and not targeted by each of the two methods. The distributions are similar---friend-based ranking achieves similar targeting performance to community-based targeting.

\begin{figure}
    \centering
    \caption{Cycle Ratio for Friend-based Ranking}
    \includegraphics[width=10cm]{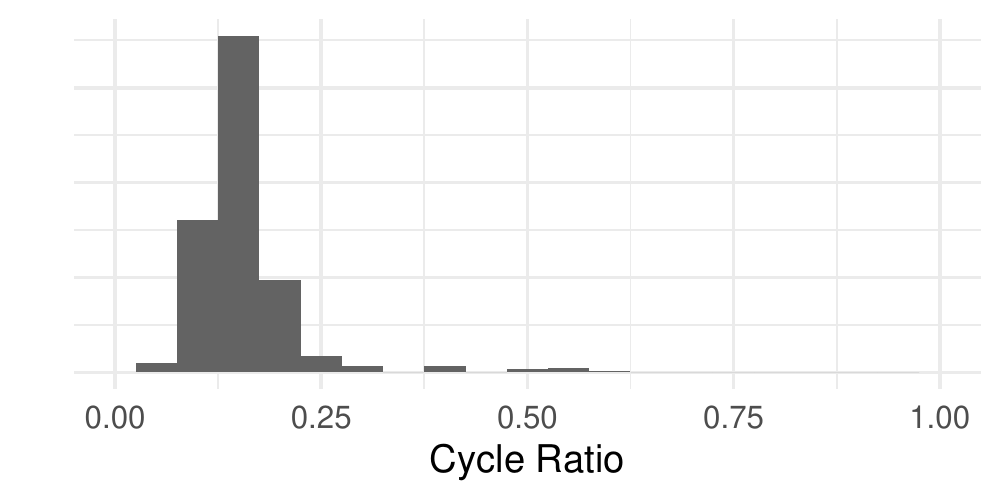}
    \label{fig:2}
    \fignote{4.5in}{Note: Sample size of 423 hamlets. The cycle ratio measures the proportion of variance in the reported ranks that can be attributed to cycles in the ranking graph.}
\end{figure}

For the friend-based ranking method, we can also report the cycle ratio for each of the 423 hamlets. Figure \ref{fig:2} shows the distribution of the cycle ratio. Most hamlets have a cycle ratio of around 0.15. Some of these cycles may derive from the weights on the edges of the ranking graph.

Although most hamlets have low cycle ratios, 10 outliers have cycle ratios of greater than 0.3. Interestingly, these 10 outliers also have low social network density---an average density of 0.25 in the outliers in comparisons to 0.7 in the remaining networks.

\bibliographystyle{aeanobold}
\bibliography{references}

@article{FBR2020,
  title={Friend-Based Ranking},
  author={Bloch, Francis and Olckers, Matthew},
  journal={American Economic Journal: Microeconomics},
  year={forthcoming}
}

@article{jiang2011hodgerank,
  title={Statistical Ranking and Combinatorial Hodge Theory},
  author={Jiang, Xiaoye and Lim, Lek-Heng and Yao, Yuan and Ye, Yinyu},
  journal={Mathematical Programming},
  volume={127},
  number={1},
  pages={203--244},
  year={2011},
  publisher={Springer}
}

@article{mattei2020,
  title={PeerNomination: Relaxing Exactness for Increased Accuracy in Peer Selection},
  author={Mattei, Nicholas and Turrini, Paolo and Zhydkov, Stanislav},
  journal={arXiv preprint arXiv:2004.14939},
  year={2020}
}

@article{alatas2012targeting,
  title={Targeting the Poor: Evidence from a Field Experiment in Indonesia},
  author={Alatas, Vivi and Banerjee, Abhijit and Hanna, Rema and Olken, Benjamin A and Tobias, Julia},
  journal={American Economic Review},
  volume={102},
  number={4},
  pages={1206--40},
  year={2012}
  }

@article{alatas2016network,
  title={Network Structure and the Aggregation of Information: Theory and Evidence from Indonesia},
  author={Alatas, Vivi and Banerjee, Abhijit and Chandrasekhar, Arun G and Hanna, Rema and Olken, Benjamin A},
  journal={American Economic Review},
  volume={106},
  number={7},
  pages={1663--1704},
  year={2016},
  publisher={American Economic Association}
  }

\end{document}